\documentclass[onefignum,onetabnum,dvipsnames]{siamonline171218}

\usepackage{lipsum}
\usepackage{caption}
\usepackage{amsfonts}
\usepackage{graphicx}
\usepackage{pythonhighlight}
\usepackage{epstopdf}
\usepackage{algorithmic}
\usepackage{tikz}

\usepackage{listings}
\usepackage{xcolor}

\definecolor{myText}{RGB}{40, 40, 40}  
\usetikzlibrary{shapes,arrows,calc}

\ifpdf
  \DeclareGraphicsExtensions{.eps,.pdf,.png,.jpg}
\else
  \DeclareGraphicsExtensions{.eps}
\fi

\usepackage{enumitem}
\setlist[enumerate]{leftmargin=.5in}
\setlist[itemize]{leftmargin=.5in}

\newsiamremark{remark}{Remark}
\newsiamremark{hypothesis}{Hypothesis}
\crefname{hypothesis}{Hypothesis}{Hypotheses}
\newsiamthm{claim}{Claim}

\headers{\texttt{optipoly}: Multi-variable polynomial box-constrained optimization}{Mazen Alamir}
\title{\texttt{optipoly}: A \texttt{Python} package for boxed-constrained multi-variable polynomial cost functions optimization}

\author{Mazen Alamir\thanks{Univ. Grenoble Alpes, CNRS, Grenoble INP, GIPSA-lab, 38000 Grenoble, France
  (\email{mazen.alamir@cnrs.fr}, \url{https://www.mazenalamir.fr}).}}

\usepackage{amsopn}

\newcommand{\e}{\ \\ \ \\}

\makeatletter
\newcommand*{\addFileDependency}[1]{
  \typeout{(#1)}
  \@addtofilelist{#1}
  \IfFileExists{#1}{}{\typeout{No file #1.}}
}
\makeatother

\newcommand*{\myexternaldocument}[1]{%
    \externaldocument{#1}%
    \addFileDependency{#1.tex}%
    \addFileDependency{#1.aux}%
}

\ifpdf
\hypersetup{
  pdftitle={Box-constrained polynomial optimization},
  pdfauthor={Mazen Alamir}
}
\fi

\myexternaldocument{ex_supplement}

\begin{document}

\maketitle

\begin{abstract}
 In this paper, a new \texttt{python} package (\texttt{optipoly}) is described that solves box-constrained optimization problem over multivariate polynomial cost functions. The principle of the algorithm is described before its performance is compared to three general purpose NLP solvers implemented in the state-of-the-art \texttt{Gekko} and \texttt{scipy} packages. The comparison show statistically better best solution provided by the algorithm with significantly less computation times. The package will be shortly made freely and easily available through the simple \texttt{pip install} process. 
\end{abstract}


\section{Introduction}
\e 
Optimizing over polynomial cost functions is an important topics in computational engineering. This is because mutli-variate polynomial function basis provide a universal representation of \textit{smooth} relationships. Moreover, optimization over multi-variate polynomial might be used as an intermediate step in a optimization process as it is the case for quadratic optimization that is commonly used in the Sequential Quadratic Programming (SQP)-based algorithm. In that sense, SQP can be viewed as a particular instance of a more general Sequential Polynomial Programming (SPP) which might be less prone to local minima as it is suggested by the comparison results shown at the end of this study. 
\e 
The starting point of the investigation that led to the algorithm involved in the \texttt{optipoly} package  is a newly developed sparse identification method, similar to the ones proposed in the \texttt{scikit-learn} library \cite{pedregosa2011scikit}, namely \texttt{larsCV}, \texttt{lassoCV} and \texttt{lassolarsCV} but which is \textbf{scalable}\footnote{Problems with 400,000 features can be solved in less than one minutes.} in the number of candidate features. This paves the way for the identification of multivariate-polynomials in a relatively high number of raw features that can represent cost functions depending on a decent number of decision variables. 
\e 
Now one might ask: \textit{What is specific about polynomial functions that makes them eligible for a special treatment that general purpose optimizers do not apply?}
\e The answer to that question lies in the fact that polynomials can benefit from very efficient vectorized computation algorithms that are implemented through the \texttt{scipy} module and its \texttt{interp1d} method. Indeed, this method enables, for instance, to evaluate a \textit{scalar} polynomial over 2 millions of values in less that 10 msec. This lies in the heart of the newly proposed algorithm that is presented in the current note. 
\e This paper is organized as follows: First of all, the principle of the algorithm is presented in Section \ref{sec-principle}. Then section \ref{sec-module} presents the \texttt{optipoly} module and its main classes and methods. Finally, Section \ref{sec-comparison} show the comparison with some state-of-the-art solvers to underline the advantages of using \texttt{optipoly}. The paper ends with Section \ref{sec_conclusion} that summarizes the message of the paper.
\section{The principle of the algorithm}\label{sec-principle}
\e 
The \texttt{optipoly} module is dedicated to solving \textbf{box-constrained polynomial} optimization problems of the following form:
\begin{equation}
\min_{x} P(x) \quad \text{$\vert\quad x\in [x_\text{min}, x_\text{max}]\subset \mathbb R^n$}\label{defduproblem}
\end{equation}
where 
\begin{itemize}
\item  $x_\text{min}, x_\text{max}\in \mathbb R^n$ stand for the vector of minimum and maximum values of the components of the decision variables $x\in \mathbb R^n$ so that the inclusion constraint expressed by:
\begin{equation}
x\in [x_\text{min}, x_\text{max}] \label{boxc}
\end{equation}
is to be interpreted component-wise. \e 
The terminology \textit{Box-constrained} refers to the fact that the constraints expressed by \eqref{boxc} are the only \textit{hard} constraints that are admissible\footnote{This implicitly means that the cost function $P(x)$ represents a polynomial approximation of some original cost to which an exact penalty on some constraints violation penalty would have been added. The approximation polynomial would incorporate all possible non box-constraints like limitations.}. \\
\item $P(x)$ is some cost function that is a multivariate polynomial in the decision variable $x\in \mathbb R^n$. More precisely, $P(x)$ takes the following general form:
\begin{equation}
P(x)= \sum_{i=1}^{n_m}{\color{RubineRed} c_i}{\color{MidnightBlue} \left[\prod_{j=1}^{n}x_j^{p_{ij}}\right]}= \sum_{i=1}^{n_m}{\color{RubineRed} c_i}{\color{MidnightBlue}\left[\phi_i(x)\right]}\label{eqpol}
\end{equation}
in which $n_m$ is the number of monomials involved in the polynomial $P$ while $\phi_i$ is a multi-variate monomial while $c_i$ is its associated coefficient in the expression of $P(x)$.
\end{itemize}
\e 
As a matter of fact, the proposed algorithm is extremely simple in that it incorporates a very old and simple principle that can be summarized as follows:
\e 
\begin{center}
\begin{tikzpicture}
\node[rounded corners, fill=gray!10, inner sep=5mm](O){
\begin{minipage}{0.8\textwidth}
\color{MidnightBlue}
Perform \textit{several}\footnote{A maximum number of iterations is defined but the iterations can be stopped before when some stopping criterion is satisfied.} successive rounds of scalar polynomial optimization-by-enumeration in each of which, one of the components of the current solution is updated. The best value of the components is updated and the whole resulting candidate vector is used to define the new scalar polynomial in the next components based on which, the next component is updated and so on.
\end{minipage}
};
\node[above] at(O.north) {\sc Principle of the algorithm};
\end{tikzpicture}
\end{center}
\e Let us say it again: \textit{Nothing is totally new here}! 
\e This principle is well known and has been \textit{rightly} abandoned based on efficiency arguments since the so called \textit{optimization-by-enumeration} is generally not efficient when standard \textit{for-loop} exploration is used. Instead descent-based iterations have been preferred since they correspond to a much lower number of function's evaluations. 
\e Notice that, if one puts aside the computation time, the enumeration process provides the a priori advantage of being less prone to local minima (as far as the scalar optimization problem is concerned) than any descent method. 
\e The efficient implementation of polynomial's evaluation over a high number of arguments values (through the \texttt{scipy-interp1d} method mentioned in the introduction) comes in handy in leveraging the advantages of enumeration while strongly reducing the computational cost associated to this process. This statement, supported by the numerical investigation presented below, is the main message of this paper.  
\e Obviously, the issue of the \textit{fight-against-the-clock} battle between descent-like approaches and cyclic scalar enumeration approaches (such as the one proposed in \texttt{optipoly}) is a quantitative question and the winner's designation is not a theoretical question. Rather, it is a practical, technological and algorithmic temporary matter.
\e Given the current situation of the latter determinant elements, the results proposed in this paper suggests that the second class of method outperforms the general purpose methods for the specific problem of box-constrained optimization of polynomial cost functions. 
\vskip 1mm
\section{Description of the \texttt{optipoly} module}\label{sec-module}
\e 
Notice first of all that the expression of the polynomial function given by \eqref{eqpol} shows that a multivariate polynomial is totally defined by the matrix of powers and the associated cost, namely:
\begin{equation}
\texttt{powers}\in \mathbb R^{n_m\times n} \quad;\quad \texttt{coefs}\in \mathbb R^{n_m}\label{powersandcoefs}
\end{equation}
which play respectively the roles of $p$ and $c$ in \eqref{eqpol}.
\e As a matter of fact, these are the only two arguments that are used to create an \textit{instance} of the \texttt{Pol} class which is the main class of the \texttt{optipoly} module. For instance, assuming that the matrix of \texttt{powers} and the vector of coefficients \texttt{coefs} are available, an instance of the \texttt{Pol} class is obtained by the following script:
\begin{center}
\vskip 3mm
\begin{tikzpicture}
\node[rounded corners, fill=Black!10, inner ysep=2mm, inner xsep=3mm]{
\begin{minipage}{0.6\textwidth}
\lstset{numbers=none}  
\begin{lstlisting}
from optipoly import Pol

powers = [[1, 0, 2], [0,3,0]]
coefs = [1.0, 2.0]

pol = Pol(powers, coefs)
\end{lstlisting}
\end{minipage}
};  
\end{tikzpicture}
\end{center} 
\vskip 2mm
The above script defines the following polynomial in the three-dimensional variable $x\in \mathbb R^3$:
\begin{equation}
P(x) = x_1x_3^2+2x_2^3 \label{defdePx}
\end{equation}
The the following script computes the values of the polynomial at the arguments defined by the lines of the following matrix $X$:
$$X:= \begin{bmatrix} 
1&1&1\cr -1&2&3\cr 0&1&0
\end{bmatrix}$$
which means that the polynomial is evaluated at the arguments:
$$\begin{bmatrix} 
1\cr 1\cr 1
\end{bmatrix}\ ,\  \begin{bmatrix} 
-1\cr 2\cr 3
\end{bmatrix}\ ,\  \begin{bmatrix} 
0\cr 1\cr 0
\end{bmatrix}$$
\begin{center}
\vskip 3mm
\begin{tikzpicture}
\node[rounded corners, fill=Black!10, inner sep=4mm]{
\begin{minipage}{0.6\textwidth}
\lstset{numbers=none}  
\begin{lstlisting}
X = [[1,1,1], [-1,2,3], [0,1,0]]
pol.eval(X)

>> array([3., 7., 2.])
\end{lstlisting}
\end{minipage}
};  
\end{tikzpicture}
\end{center} 
Before the attributes and the methods of the class are described, let us mention the methods that enable, at a given value of $x$ to extract the scalar polynomial in a specific in terms of a specific component of the decision variable. 
\e 
In the following script, we use the \texttt{extract\_sc\_pol} method of the \texttt{pol} instance to compute the \texttt{scipy} scalar polynomial in $x_2$ at the current argument $x_0=(-1,2,3)$ which obviously leads to the polynomial given by 
\begin{equation}
2x_2^3-9 \label{lepolatx0}
\end{equation}
which is identical to the result of the script.
\begin{center}
\vskip 3mm
\begin{tikzpicture}
\node[rounded corners, fill=Black!10, inner sep=4mm]{
\begin{minipage}{0.8\textwidth}
\lstset{numbers=none}  
\begin{lstlisting}
x0 = np.array([-1,2,3])
sc_pol = pol.extract_sc_pol(x=x0, ix=1)
sc_pol

>> poly1d([ 2.,  0.,  0., -9.])
\end{lstlisting}
\end{minipage}
};  
\end{tikzpicture}
\end{center}
\e 
The following sections list the attributes of an instance of the class \texttt{Pol}, its methods and some useful methods of the class \texttt{Pol} itself that might be conveniently used for different purposes.
\subsection{The attributes of the instances of the \texttt{Pol} class}
\e 
Once the instance \texttt{pol} of the class \texttt{Pol} is created, it exports the attributes that are listed in Table \ref{table_attributes}. These are shortly given in the following list:
$$ \texttt{powers}, \quad\texttt{coefs}\quad,\texttt{nx},\quad\texttt{ncoefs}\quad\texttt{deg}$$
which are rather self-explanatory.
\captionsetup[table]{skip=10pt, justification=centering}
\begin{table}[H]
    \centering
    \begin{tabular}{|l|p{0.8\textwidth}|}
        \hline
        \textbf{Attribute} & \textbf{Definition} \\
        \hline\hline  
        \texttt{\color{MidnightBlue} powers} & The matrix of powers used in the instantiation \\
        \hline
        \texttt{\color{MidnightBlue} coefs} & The vector of coefficients of monomials used in the instantiation. \\
        \hline
        \texttt{\color{MidnightBlue} nx} & The number of variables (dimension of $x$) \\
        \hline
        \texttt{\color{MidnightBlue} ncoefs} & The number of monomials ($n_m$ in \eqref{eqpol}) \\
        \hline
        \texttt{\color{MidnightBlue} deg} & The degree of the multi-variate polynomial (the maximum values of the sum of powers over the lines of \texttt{powers}. \\
        \hline 
    \end{tabular}
    \caption{List of the attributes of an instance of the class \texttt{Pol} of the \texttt{optipoly} module.}\label{table_attributes}
\end{table}
\subsection{The methods exposed by an instance of the class \texttt{Pol}}
\e These methods are listed in Table \ref{table_instance_meth}. As an example, the use of the \texttt{to\_df} method on the previously presented sample instance of the class \texttt{Pol} leads to the following dataframe 
\begin{center}
\vskip 3mm
\begin{tikzpicture}
\node[rounded corners, fill=Black!10, inner sep=4mm]{
\begin{minipage}{0.25\textwidth}
\lstset{numbers=none}  
\begin{lstlisting}
df = pol.to_df()
df
\end{lstlisting}
\end{minipage}
};  
\end{tikzpicture}
\includegraphics[width=0.3\textwidth]{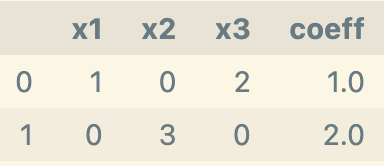} 
\end{center}
\e 
As for the \texttt{to\_dict} method, it results in the following dictionary:
\begin{center}
\vskip 3mm
\begin{tikzpicture}
\node[rounded corners, fill=Black!10, inner sep=4mm]{
\begin{minipage}{0.4\textwidth}
\lstset{numbers=none}  
\begin{lstlisting}
dic = pol.to_dict()
dic
\end{lstlisting}
\end{minipage}
};  
\end{tikzpicture}
\includegraphics[width=0.35\textwidth]{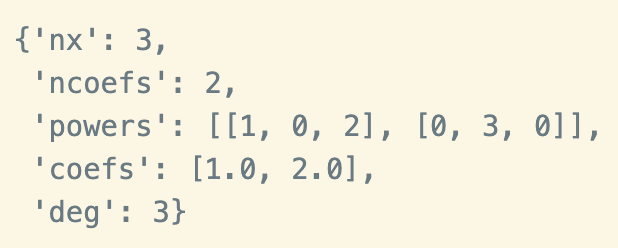} 
\end{center} 
\e Since the definition of the call for the main method \texttt{solve} is more involved, the list of arguments and their description is given in the next section. 
\begin{table}[H]
    \centering
    \begin{tabular}{|l|p{0.13\textwidth}|p{0.13\textwidth}|p{0.42\textwidth}|}
        \hline
        \textbf{Methods} & \textbf{input} & \textbf{returns} & \textbf{Description} \\
        \hline\hline  
        \texttt{\color{MidnightBlue} extract\_sc\_pol} & \texttt{x0}, \texttt{ix} & \texttt{sc\_pol} & Generate a \texttt{scipy} scalar polynomial \texttt{sc\_pol} in the component indexed by \texttt{ix} at the current value \texttt{x0} of the decision variable.\\
        \hline 
        \texttt{\color{MidnightBlue} to\_df} & \texttt{None}& \texttt{df} & Generate a \texttt{pandas} dataframe \texttt{df} representing the polynomial data.\\
        \hline
        \texttt{\color{MidnightBlue} to\_dict} & \texttt{None}& \texttt{dic} & Generate a \texttt{python} dictionary \texttt{dic} representing all the information regarding the polynomial.\\
        \hline 
        \texttt{\color{MidnightBlue} solve} & \texttt{x0}, \texttt{xmin}, \texttt{xmax}, \texttt{Ntrials}, \texttt{ngrid}, \texttt{iter\_max}, \texttt{eps}, \texttt{psi}& \texttt{solution}, \texttt{cpu} & The main method that minimizes/maximizes or find roots for the instance polynomial. The mode depends on the choice of the \texttt{lambda} function \texttt{psi}. See the detailed description of the parameters in Section \ref{secsolve}.\\
        \hline 
    \end{tabular}
    \caption{List of methods exported by an instance of the class \texttt{Pol} of the \texttt{optipoly} module.}\label{table_instance_meth}
\end{table}
\subsection{Description of the main \texttt{solve} method}\label{secsolve}
The last method (\texttt{solve}) invoked in Table \ref{table_instance_meth} which is the \textit{raison d'être} of the \texttt{optiply} package is more extensively described in the present section. Let us start by the discussion of the inputs arguments: 
\subsubsection{The input arguments for the \texttt{solve} method}
The input arguments of the \texttt{solve} method are described in Table \ref{table_solve_meth}.
\begin{table}[H]
    \centering
    \begin{tabular}{|l|p{0.7\textwidth}|}
        \hline
        \textbf{Parameters} & \textbf{Description}\\
        \hline\hline  
        \texttt{\color{MidnightBlue} x0} &  The initial guess for the solver. This is a vector of dimension \texttt{nx}. Notice that when several starting points are used (\texttt{Ntrials}$>1$ as explained below), the next initial guesses are randomly sampled in the admissible domain defined by \texttt{xmin} and \texttt{xmax}.\\
        \hline
        \texttt{\color{MidnightBlue} xmin} &  The vector of lower bounds of the decision variable. \\ 
        \hline 
        \texttt{\color{MidnightBlue} xmax} & The vector of upper bounds of the decision variable.\\
        \hline
        \texttt{\color{MidnightBlue} Ntrials} & The number of different starting points used in order to enhance the avoidance of local minima (default is set to 1). \\
        \hline 
        \texttt{\color{MidnightBlue} ngrid} & The number of grid points used in the scalar optimization-by-enumeration in the different direction of the components of the decision variable (default is set to 1000). \\
        \hline
        \texttt{\color{MidnightBlue} iter\_max} & Maximum number of rounds of scalar optimization (default is set to 100). This number is rarely used since the optimization is stopped before based on the \texttt{eps}-related termination condition.  In almost all the 1000 problems used in the benchmark discussed in this paper the number of iterations never exceeded 3 or 4.\\
        \hline
        \texttt{\color{MidnightBlue} eps} & 
        The precision that is used to stop the iterations (default value set to $10^{-2}$). More precisely, the iterations stop when the last round of scalar optimization does not improve the cost function by more than $$\texttt{eps}\times\vert\texttt{J\_{previous}}\vert$$
        where \texttt{J\_{previous}} is the cost achieved at the previous round. \\
        \hline
        \texttt{\color{MidnightBlue} psi}  & The lambda function that applies to the cost function in order to define the modified quantity that is to be minimized. For instance the choice:
\begin{center}
\texttt{psi}= \texttt{lambda} v : -v
\end{center}
leads to the \texttt{solver} method being oriented towards the maximization of the original cost. On the other hand, the choice:
\begin{center}
\texttt{psi}= \texttt{lambda} v : abs(v)
\end{center}
leads to the \texttt{solver} method being oriented towards finding a root of the polynomial. The default setting is given by 
\begin{center}
\texttt{psi}= \texttt{lambda} v : v
\end{center}
leading to \texttt{solve} trying to minimize the cost function.
Notice that the above are only possible choices but any expression of $v$ might be used provided that it is compatible with vectorized evaluation.
\ \\
\hline
    \end{tabular}
    \caption{Description of the input parameters of the \texttt{solve} method of the \texttt{Pol} class.}\label{table_solve_meth}
\end{table}
\subsubsection{The output arguments of the \texttt{solve} method}
The call for the \texttt{solve} method of an instance \texttt{pol} of the class \texttt{Pol} takes the following syntax:
\begin{center}
\vskip 3mm
\begin{tikzpicture}
\node[rounded corners, fill=Black!10, inner sep=4mm]{
\begin{minipage}{0.8\textwidth}
\lstset{numbers=none}  
\begin{lstlisting}
solution, cpu = pol.solve(...)
\end{lstlisting}
\end{minipage}
};  
\end{tikzpicture}
\end{center}
\vskip 2mm where \texttt{solution} is a dictionary with the fields \texttt{x} and \texttt{f} standing respectively for the best found solution and the associated best corresponding value. On the other hand, \texttt{cpu} is the computation time required to produce the results. 
\e The following script gives an example of a call that asks for the maximization of the polynomial defined earlier in the paper then prints the results so obtained: 
\begin{center}
\vskip 3mm
\begin{tikzpicture}
\node[rounded corners, fill=Black!10, inner sep=4mm]{
\begin{minipage}{0.8\textwidth}
\lstset{numbers=none}  
\begin{lstlisting}
nx = 3
x0 = np.zeros(nx)
ntrials = 6
ngrid = 1000
xmin = -1*np.ones(nx)
xmax = 2*np.ones(nx)

solution, cpu = pol.solve(x0=x0, 
                          xmin=xmin, 
                          xmax=xmax, 
                          ngrid=ngrid, 
                          Ntrials=ntrials, 
                          psi=lambda v:-v
                          )
                          
print(f'xopt = {solution.x}')
print(f'fopt = {solution.f}')
print(f'computation time = {solution.cpu}')
\end{lstlisting}
\end{minipage}
};  
\end{tikzpicture}
\end{center} 
\e 
This call yields the following results:
\begin{center}
\vskip 3mm
\begin{tikzpicture}
\node[rounded corners, fill=Black!10, inner sep=4mm]{
\begin{minipage}{0.8\textwidth}
\lstset{numbers=none}  
\begin{lstlisting}
xopt = [-1.  2.  0.]
fopt = 16.0
computation time = 0.0046999454498291016
\end{lstlisting}
\end{minipage}
};  
\end{tikzpicture}
\end{center}
\e Changing the definition of \texttt{psi} to 
\texttt{\color{MidnightBlue} psi=lambda v:abs(v)} leads to the following results:
\begin{center}
\vskip 3mm
\begin{tikzpicture}
\node[rounded corners, fill=Black!10, inner sep=4mm]{
\begin{minipage}{0.8\textwidth}
\lstset{numbers=none}  
\begin{lstlisting}
xopt = [-0.996997    0.58858859  0.63963964]
fopt = -9.305087356087371e-05
computation time = 0.003011941909790039
\end{lstlisting}
\end{minipage}
};  
\end{tikzpicture}
\end{center}
\e Finally using the default setting \texttt{\color{MidnightBlue} psi=lambda v:v} makes the method focused on finding the minimum of the cost function which leads to the following results:
\begin{center}
\vskip 3mm
\begin{tikzpicture}
\node[rounded corners, fill=Black!10, inner sep=4mm]{
\begin{minipage}{0.8\textwidth}
\lstset{numbers=none}  
\begin{lstlisting}
xopt = [-1. -1.  2.]
fopt = -6.0
computation time = 0.005150318145751953
\end{lstlisting}
\end{minipage}
};  
\end{tikzpicture}
\end{center}
\subsection{The \texttt{Pol} class's method}
\e 
Beside the previously described attributes and methods that are exported by the instances of the \texttt{Pol} class. The following method is a  \textit{class methods} that does not depend on a specific instance of the \texttt{Pol} class:
\e 
\texttt{\color{MidnightBlue} generate\_random\_pol}\e This method accepts the following input arguments:
\begin{tabbing}
\hskip 2cm \= \hskip 5cm\\
\texttt{\color{MidnightBlue} nx}\> The number of variables of the polynomial. \\
\texttt{\color{MidnightBlue} deg\_max}\> The degree of the polynomial.\\
\texttt{\color{MidnightBlue} card}\> The number of monomials involved in the polynomial's expression.
\end{tabbing}
\e 
Based on the above input arguments, the method returns an instance of the class \texttt{Pol} that is a polynomial in \texttt{nx} variables, of degree \texttt{deg\_max} involving \texttt{card} monomials. The choice of the matrix of powers and the vector of coefficients needed to instantiate the class are randomly sampled.
\e 
\section{Comparison with some state-of-the-art general purpose solvers}\label{sec-comparison}
\e 
In this section, the performance of the above discussed algorithm in terms of the quality of the result found on one hand and the required computation time on the other hand are examined. 
\e 
First of all, the alternative solvers are listed and shortly described, then the definition of the benchmark before the results are given.
\subsection{The alternative solvers}
\e
The set of alternative algorithms considered in the following comparison includes two NLP solvers called via the \texttt{Gekko} modeling framework \cite{beal2018gekko} and a global optimization oriented solver proposed by the famous \texttt{scipy} python module. \e 
More precisely, the following solvers are considered:\\
\begin{itemize}
\item[$\checkmark$] \textbf{\color{MidnightBlue} IPOPT}: The famous interior point NLP solver \cite{biegler2009large}.\\
\item[$\checkmark$] \textbf{\color{MidnightBlue} BPOPT} Another NLP general-purpose solver that is made available through the \texttt{Gekko} framework\footnote{Notice that there is a third NLP algorithm proposed via the \texttt{Gekko} framework which is called APOPT that seems to produce exactly the same results than the \texttt{IPOPT} solver over the 1000 problem's instances which seems weird. That is the reason whyy this third algorithm has not be included in the comparison.}.\\
\item[$\checkmark$] \textbf{\color{MidnightBlue} Dual-Annealing}: A \texttt{scipy} general-purpose solver for \textit{global optimization} \cite{1996Tsallis, 2020SciPy}. This class of algorithm is interesting to consider here since the issue of local minima is one of the key issues to investigate since high degree multivariate polynomial generally present multiple local minima as it is confirmed by the results shown later on in this paper.
\end{itemize}
\subsection{The benchmark's definition}
\e 
The benchmark consists in 1000 experiments of box-constrained polynomial optimization problems. For each experiment, a randomly sampled polynomial is generated using values of \texttt{nx}, \texttt{deg} and \texttt{card} that span the sets of values produced by the following script\footnote{Notice that the product of the cardinalities of the sets is equal to $1000$.}:
\begin{center}
\vskip 3mm
\begin{tikzpicture}
\node[rounded corners, fill=Black!10, inner sep=4mm]{
\begin{minipage}{0.9\textwidth}
\lstset{numbers=none}  
\begin{lstlisting}
set_nx = [3, 5, 10, 20, 50]
set_deg = [int(i) for i in list(np.arange(2,10))]
set_card = [int(i) for i in list(np.arange(5,30))]
configurations = list(itertools.product(set_nx, set_deg, set_card))
list_of_pols = [generate_random_pol(*c) for c in configurations]
\end{lstlisting}
\end{minipage}
};  
\end{tikzpicture}
\end{center} 
\e 
This results in a list, called \texttt{configurations} in which each element is a triplet of values of (\texttt{nx}, \texttt{deg}, \texttt{card}) that can be used as calling list of arguments to the \texttt{generate\_random\_pol} described above.
\newpage 
\subsection{Comparison results}\label{subsec-comparison}
\begin{figure}[H]
    \centering
    \framebox{\includegraphics[width=0.95\linewidth]{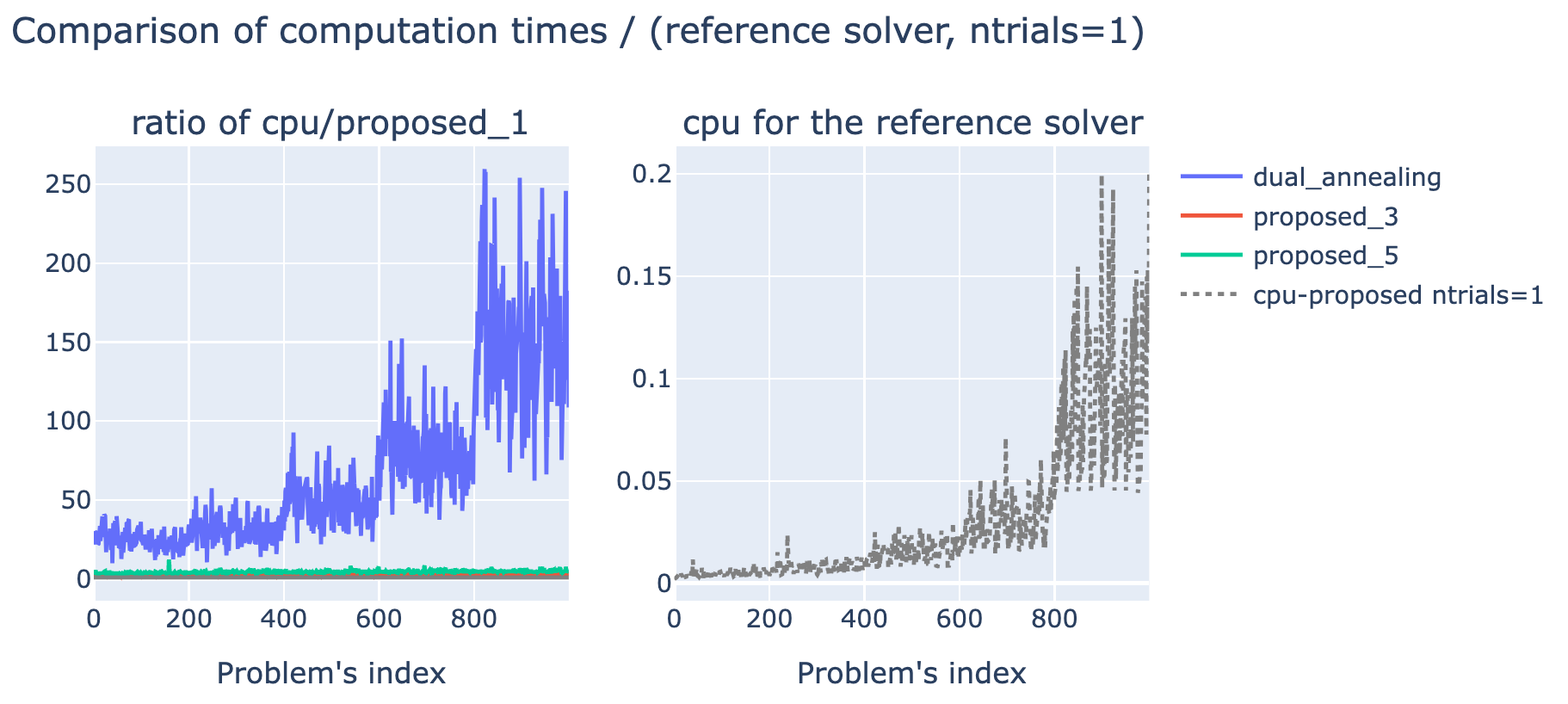}}
    \caption{Histogram of the ratios of computation time relatively to the proposed one using \texttt{Ntrials=1}.(Left) when a curve is above the horizontal gray line, this means that the cpu time for the corresponding alternative solver is greater than the reference cpu time of the proposed algorithm with a single trial \texttt{Ntrials}=1. (Right) The reference cpu time of the latter algorithm.}
    \label{fig_comparing_cpus_proposed}
\end{figure}

\ \e
Two aspects of the performance are investigated in this section, namely:\e 
\begin{itemize}
\item \textbf{\color{MidnightBlue} 
The best found values of the cost function}. 
\e More precisely, in order to compare the performance of the proposed algorithm to the three above mentioned ones, the following differences are examined:
\begin{equation}
 \Delta_h := f_h-f_\text{proposed}\vert_{\text{Ntrials}=1}\quad h\in \{\text{IPOPT, BPOPT, Dual-Annealing}\}\label{defdeDelt_h}
\end{equation} 
where $f_h$ is the best value of the cost function that is achieved by the algorithm $h$ while $f_\text{proposed}\vert_{\text{Ntrials}=1}$ is the value achieved by the proposed algrothm using only one trial (\texttt{Ntrials}=1).\e Obviously, positive value of $\Delta_h$ means that the proposed algorithm achieved better solution that the alternative algorithm $h\in \{\text{IPOPT, Dual-Annealing, BPOPT}\}$.
\e The results are shown in Figure \ref{fig_comparing_values} where the histograms of $\Delta_h$ is shown suggesting that in the majority of the problems, the proposed solver outperforms the alternative \texttt{gekko} \textit{non global optimization-oriented} solvers (this is strictly true all the time when \texttt{bpopt} is considered). 
\e In the case of the IPOPT solver, the proposed solver  shows better results for a large majority of problems. Moreover, when this does not happens, the difference is less important that the ones showing the superiority of the proposed algorithm. This can be more clearly seen on the cumulative histogram shown in Figure \ref{fig_comparing_values_2}. \e 
As for the global-optimization oriented solver, it provides better results in on a slightly more frequent set of problems but the proposed algorithm still outperforms this global algorithm on a non negligible number of instances. 
\e Notice however that the global-optimizer needs computation times that are largely greater than the ones needed by the proposed algorithm as it is shown below (see Figure \ref{fig_comparing_cpus_proposed}. Moreover, by using \texttt{Ntrials}=3 or 5, the proposed solver systematically outperforms the global \texttt{scipy} optimizer as it is shown in Figure \ref{fig_compare_values_proposed} while keeping lower computation times as shown in Figure \ref{fig_comparing_cpus_proposed}.
\begin{figure}
    \centering
    \framebox{\includegraphics[width=0.85\linewidth]{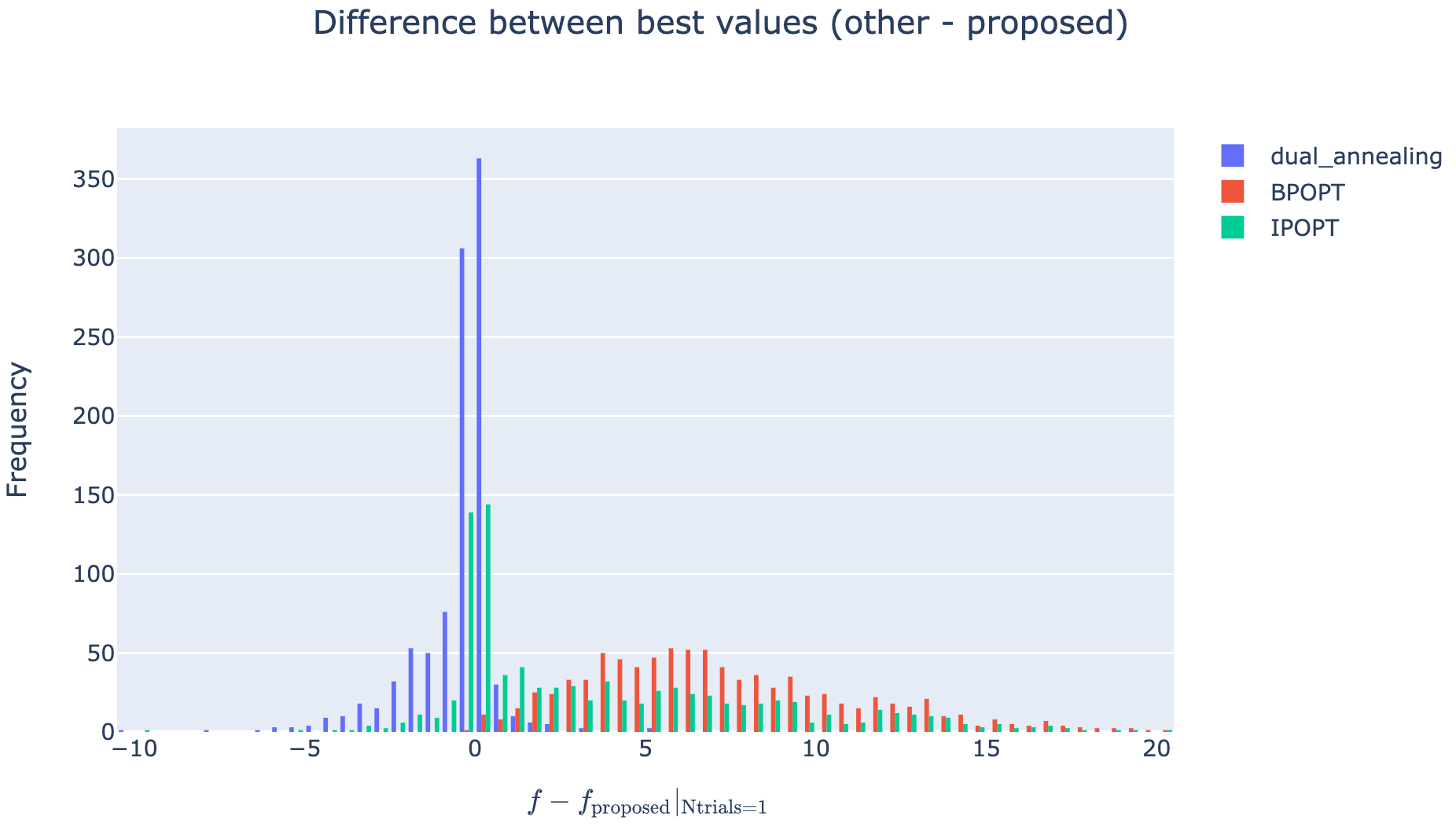}}
    \caption{Histogram of the difference $f-f_\text{proposed}\vert_\text{Ntrials=1}$. A positive difference means that the best value obtained by the alternative solution is higher than the one delivered by the proposed reference solver using a single trial \texttt{Ntrials}=1. Although the global solver shows statistically better results, its computation times is largely higher as shown in Figure \ref{fig_comparing_cpus}.}
    \label{fig_comparing_values}
\end{figure}
\e 
\begin{figure}[H]
    \centering
    \framebox{\includegraphics[width=0.95\linewidth]{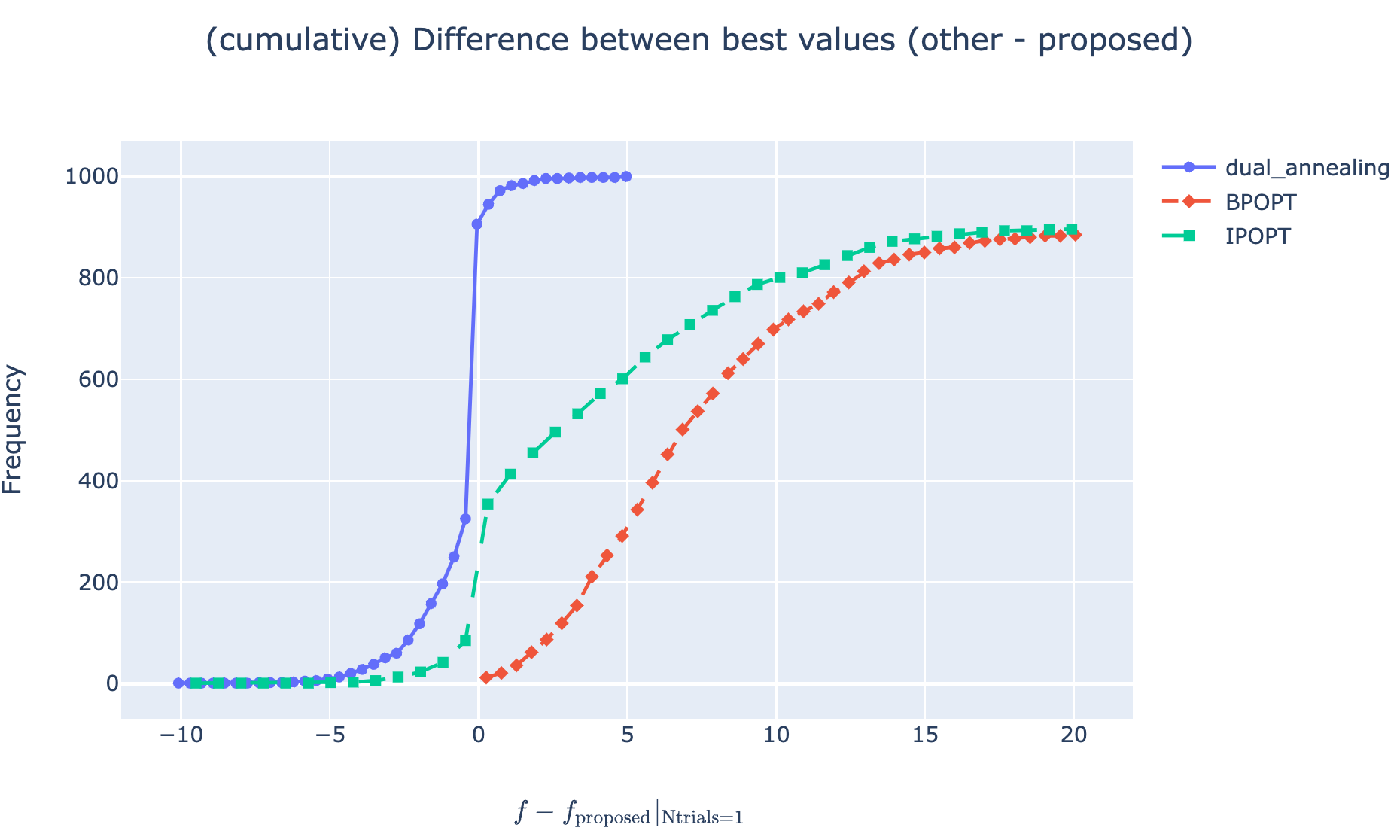}}
    \caption{Cumulative version of the histogram of Figure \ref{fig_comparing_values}. This figure shows also that the \texttt{gekko} solvers were not able to solve some of the problems considered (symbolic relationships were too long!)}
    \label{fig_comparing_values_2}
\end{figure}
\begin{figure}[H]
    \centering
    (a)\e 
    \includegraphics[width=0.95\linewidth]{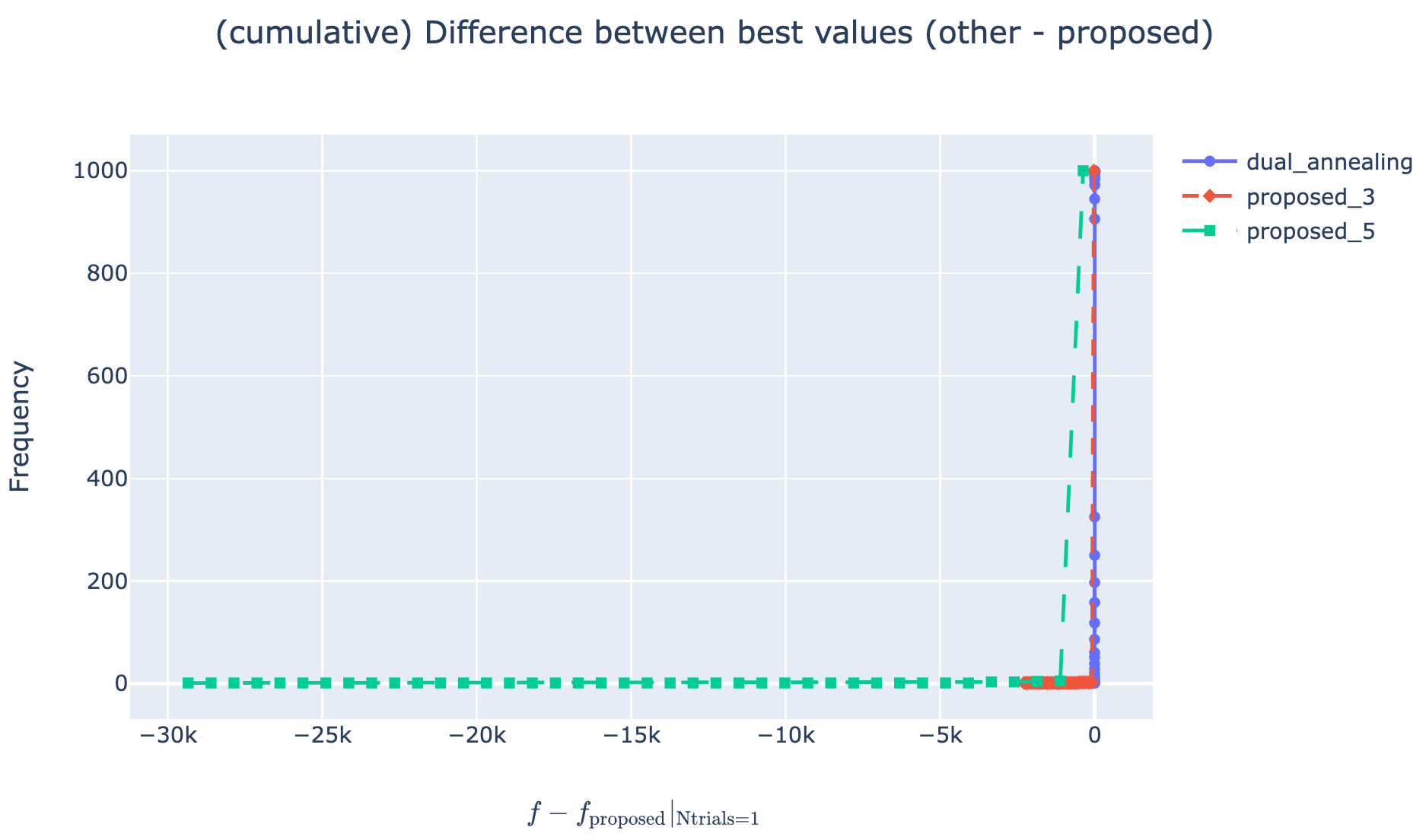} \ \\
    (b) \e 
    \includegraphics[width=0.95\linewidth]{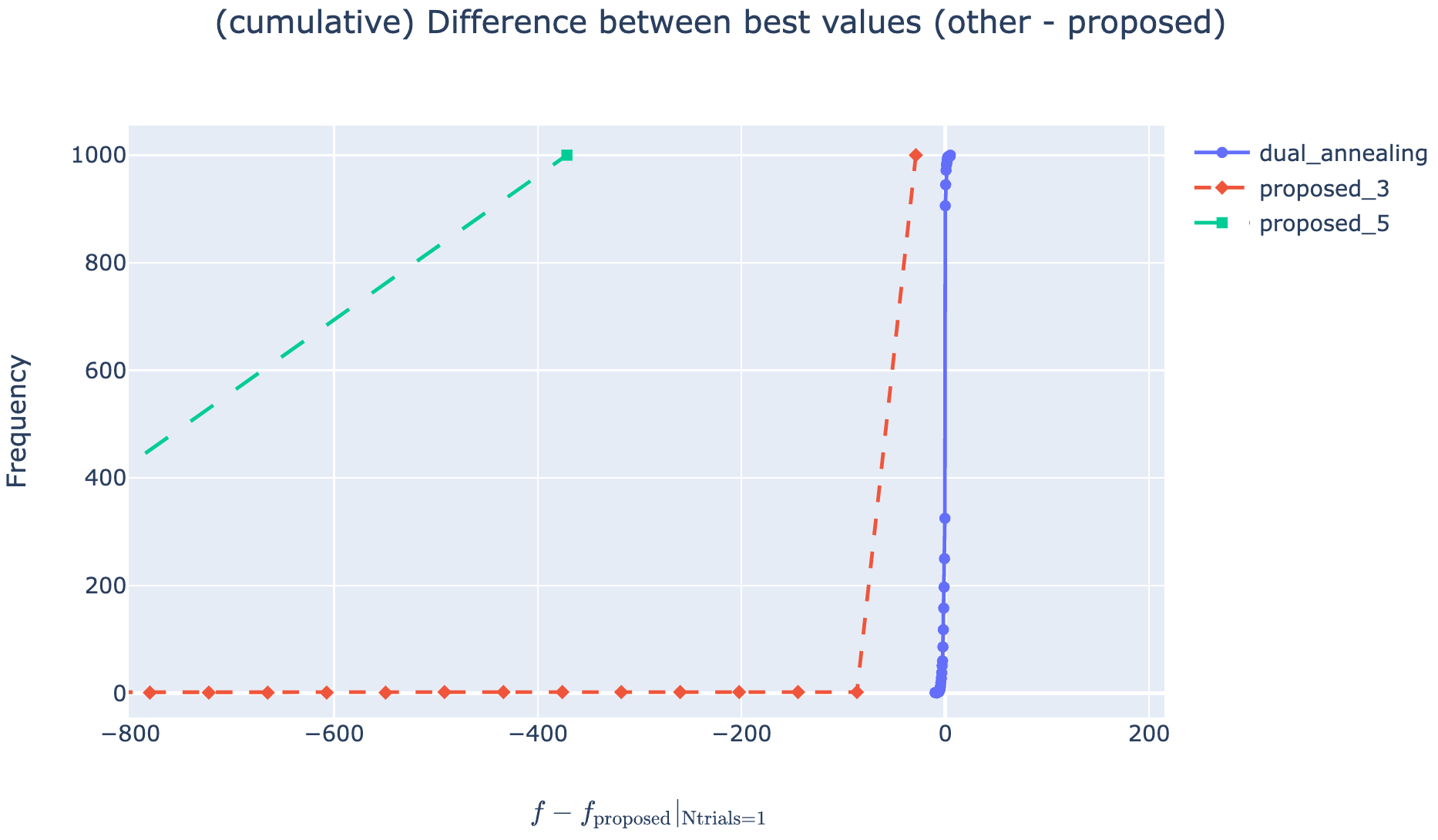}
    \caption{(a) Cumulative histograms of achieved best values (relatively to the reference solver with \texttt{Ntrials}=1) for the \texttt{dual\_annealing} \texttt{scipy} optimizer and the proposed algorithm with \texttt{Ntrials}=3 and 5. (b) zoom on the plots of (a) for a better reading of the curves close to $0$.}
    \label{fig_compare_values_proposed}
\end{figure}
\e The examination of Figure \ref{fig_comparing_values_2} showing the cumulative histogram of the incremental difference in the best achieved values compared to the reference solver (the proposed one with \texttt{Ntrials}=1) shows that the solvers implemented through the \texttt{Gekko} framework were not able to solve all the problems included in the benchmark. The error message mentioned some difficulties that were due the maximum length of the symbolic relationship involved in the modeling process. This happens for high values of \texttt{nx}=50.
\e
This being said, it is not the message of this paper to underline this shortcoming since the author is not sure whether this problem could have been by-passed by some specific manipulations (some are proposed in the warning messages). 
\e Notice that no such remedy was searched for because the resulting qualitative conclusion would remained valid given the results already obtained on the large set of problems on which the \texttt{Gekko}-implemented solver were successful in providing solutions. \e 
\item \textbf{\color{MidnightBlue} The computation time}. All the computations are performed on a \texttt{Apple M3 Pro, 18 Go}. The default settings of the alternative solvers implemented in \texttt{Gekko} is used with the following two options:
\begin{center}
\vskip 3mm
\begin{tikzpicture}
\node[rounded corners, fill=Black!10, inner sep=4mm]{
\begin{minipage}{0.35\textwidth}
\lstset{numbers=none}  
\begin{lstlisting}
options.MAX_ITER = 5000
options.OTOL = 1e-9
\end{lstlisting}
\end{minipage}
};  
\end{tikzpicture}
\end{center}
\e 
The computation times are reported in Figure \ref{fig_comparing_cpus} where the left figure shows the ratios between the computation time of the alternative solvers and the computation time required by the proposed algorithm with \texttt{ntrials}=1. When the plots are above the horizontal gray line (corresponding to 1), this means that the computation time used by the alternative solvers is greater than the on needed by the reference proposed solver with \texttt{Ntrials}=1. 
\e On the right subplot of Figure \ref{fig_comparing_cpus}, the computation time for the proposed solver is given for all the experiments involving the reference proposed solver. Moreover Figure \ref{fig_comparing_cpus_proposed} shows the same results except that the time needed by the proposed solver with different settings of the parameter \texttt{Ntrials}=3 and 5 are shown in order to show that these last settings enables the solver to largely outperform the global \texttt{scipy} optimizer while keeping largely lower computation times. 
\end{itemize}
\begin{figure}[H]
    \centering
    \framebox{\includegraphics[width=0.95\linewidth]{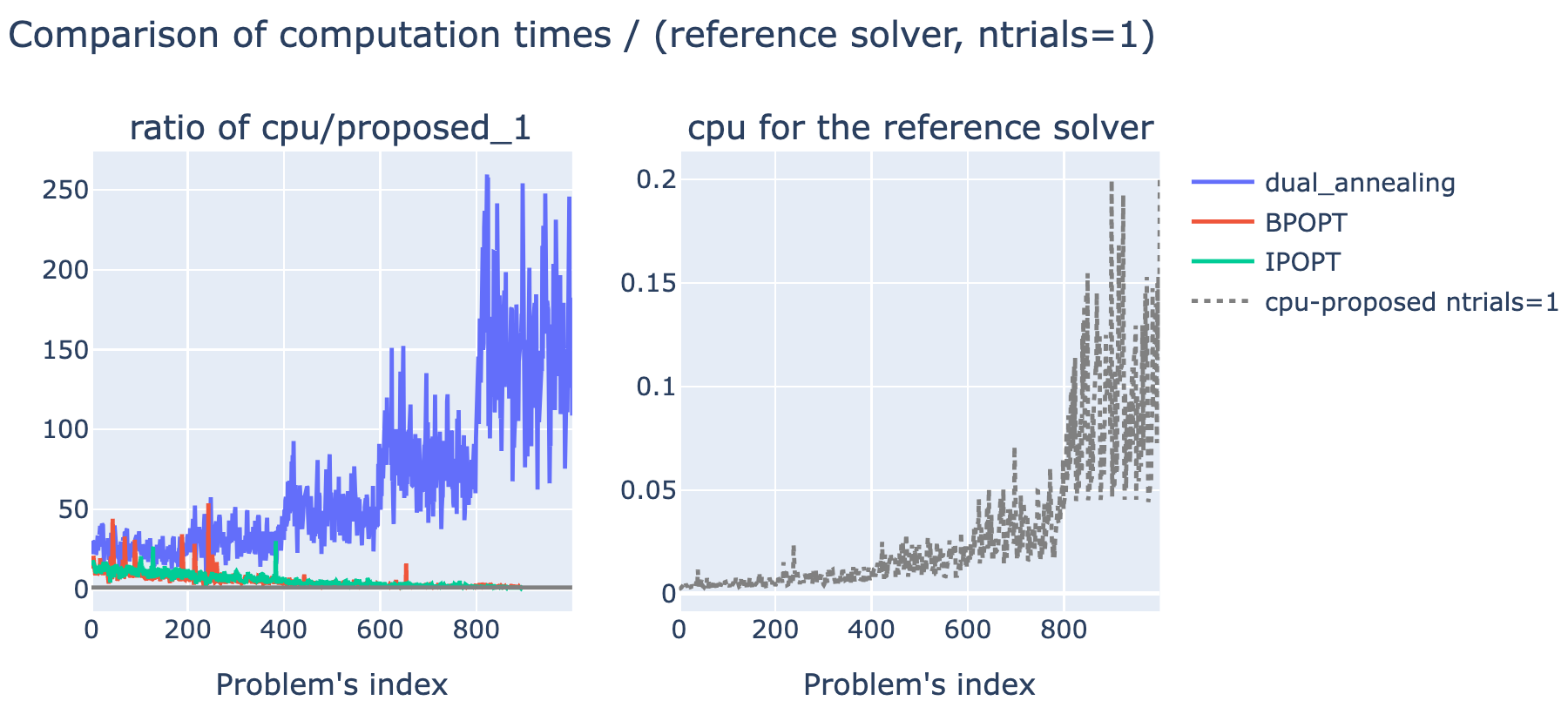}}
    \caption{Histogram of the ratios of computation time relatively to the proposed one using \texttt{Ntrials=1}.(Left) when a curve is above the horizontal gray line, this means that the cpu time for the corresponding alternative solver is greater than the reference cpu time of the proposed algorithm with a single trial \texttt{Ntrials}=1. (Right) The reference cpu time of the latter algorithm.}
    \label{fig_comparing_cpus}
\end{figure}
These results coupled with the comparative performance results shown in Figures \ref{fig_comparing_values}, \ref{fig_comparing_values_2} and \ref{fig_compare_values_proposed} clearly suggest that the proposed solver, even with a single trial, provide better performance with less computation time. 
\begin{figure}[H]
    \centering
\framebox{\includegraphics[width=0.8\linewidth]{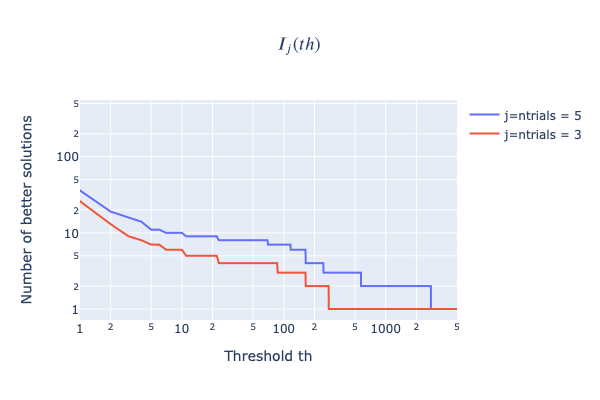}}
    \caption{The number of problem's instances for which the use of \texttt{Ntrials}=3 or 5 provides better results in the sense of \eqref{th} depending on the threshold \texttt{th} being used.}
    \label{fig_ntrials}
\end{figure}
\e 
The last investigated aspect concerns the relevance of using multiple starting points in the proposed algorithm which can be performed using higher values of the settings parameter \texttt{Ntrials}. 
\e 
Recall that the enumeration aspect over each single direction is already providing a sort of partial immunity against local minima. 
\e Nevertheless, there is no guarantee that even when no improvement is possible to obtain by changing a single direction alone, this does not mean that the current iterate is a local minimum since it would be possible to get improvement by moving more than a single component at once. 
\e This is the \textit{raison d'être} of the multiple-starting-point mechanism associated with the use of \texttt{Ntrials}$>1$.
\e In order to examine this aspect, the same set of 1000 experiments have been also solved using the proposed algorithm with the two following values of the \texttt{Ntrials} argument:
$$\texttt{Ntrials}\in \{3, 5\}$$
and the number of experiments for which the use of multiple starting point has improved the result by more than a threshold \texttt{th} has been reported. More precisely, the cardinality of the following \texttt{th}-dependent set:
\begin{equation}
I_j(\texttt{th}) := \text{card}\Big\{\text{experiment}\ \vert\ \dfrac{f_\text{proposed}\vert_{\text{Ntrials}=j}-f_\text{proposed}\vert_{\text{Ntrials}=1}}{\vert f_\text{proposed}\vert_{\text{Ntrials}=1}\vert } \le -\texttt{th}\Bigr\} \quad j\in \{3,5\}\label{th}
\end{equation}
The results are shown in Figure \ref{fig_ntrials}. 
\section{Conclusion}\label{sec_conclusion}
\e 
In this paper, a \texttt{python} module called \texttt{optipoly} incorporating a new simple algorithm is proposed for the minimization of polynomial cost function.
\e The algorithm exploits the \texttt{python-interp1d} powerful computation in order to perform iterative scalar optimization along single components that are circularly updated. The algorithm is embedded inside a \texttt{Pol}ynomial class that exports a \texttt{solve} method that enables to run the underlying algorithm. 
\e The comparison with some state-of-the-art freely available algorithms shows that, as far as polynomial cost is concerned, the proposed algorithm outperforms both standard non-global solvers as well as global solvers in terms of computation times as well as in terms of the quality of the best solution found. 
\e The corresponding python module \texttt{polyopt} will be made available soon through the \texttt{pip} installation process: 
\begin{center}
\begin{center}
\vskip 3mm
\begin{tikzpicture}
\node[rounded corners, fill=Black!10, inner sep=4mm]{
\begin{minipage}{0.8\textwidth}
\lstset{numbers=none}  
\begin{lstlisting}
pip install optipoly
\end{lstlisting}
\end{minipage}
};  
\end{tikzpicture}
\end{center}
\end{center}
\ \\
Following the syntax of use mentioned throughout the paper. 
\newpage
\bibliographystyle{siamplain}
\bibliography{references}

\end{document}